# Synthesis of possible room temperature superconductor LK-99: $Pb_9Cu(PO_4)_6O$

Kapil Kumar[1,2], N.K. Karn[1,2], and V.P.S. Awana[1,2]

[1]*CSIR-National Physical Laboratory, Dr. K. S. Krishnan Marg, New Delhi-110012, India.*
[2]*Academy of Scientific and Innovative Research (AcSIR), Ghaziabad 201002, India.*

**Abstract:**

The quest for room temperature superconductors have been teasing the scientists and physicists, since its inception in 1911 itself. Several assertions have already been made about room temperature superconductivity, but were never verified or reproduced across the labs. The cuprates were the earliest high transition temperature ($T_c$) superconductors, and it seems that copper has done the magic once again. Last week, a Korean group synthesized a Lead Apatite based compound LK-99, showing a $T_c$ of above 400K [1-3]. The signatures of superconductivity in the compound are very promising, in terms of resistivity ($\rho = 0$) and diamagnetism at $T_c$. Although, the heat capacity ($C_p$) did not show the obvious transition at $T_c$. Inspired by the interesting claims of above room temperature superconductivity in LK-99, in this article, we report the synthesis of polycrystalline samples of LK-99, by following the same heat treatment as reported in [1,2] by two step precursor method. The phase is confirmed through X-ray diffraction (XRD) measurements, performed after each heat treatment. The room temperature diamagnetism is not evidenced by the levitation of a permanent magnet over the sample or vice versa. Further measurements for the confirmation of bulk superconductivity on variously synthesized samples are underway. Our results on present LK-99 sample, being synthesized at $925^0$C, as of now do not approve the appearance of bulk superconductivity at room temperature. Further studies with different heat treatments are though, yet underway.

***Corresponding Author**
Dr. V. P. S. Awana: E-mail: awana@nplindia.org
Ph. +91-11-45609357, Fax-+91-11-45609310
Homepage: awanavps.webs.com

## Introduction:

Superconductivity is one of the most celebrated phenomena in the history of condensed matter physics. The observation of room-temperature superconductivity has been very fascinating, as it is supposed to make huge leap in technology and associated condensed matter physics. A great deal of studies has gone into creating a material that may exhibit superconductivity at ambient conditions. Although room temperature superconductivity has been achieved, but it requires massive pressure (several GPa) making them technologically inapplicable [4-10]. Recently, few studies have shown the hope of achieving superconductivity at ambient conditions i.e., at room temperature and atmospheric pressure [1-3 and 11], but their independent confirmations by other experimental groups is yet warranted. The list includes the Ag/Au nanoparticles [12] and more recently the Nitrogen-doped Lutetium Hydride [13]. These experiments gathered the significant interest of condensed matter scientists, but their efforts went in vain as the claims could not be reproduced [12-14].

In this perspective, very recently (25$^{th}$ July 2023), room-temperature superconductivity is achieved by doping Copper in a known compound Lead Apatite $Pb_{10}(PO_4)_6O$ [1,2]. The copper doped Lead Apatite is named as LK-99. The claims are well supported by the experimental data as the superconducting phase is observed in both magnetic and resistivity measurements, fascinatingly the magnetic levitation video is also reported [2,15]. The authors also provided possible explanation of superconductivity embedded in the structure of LK-99. The doping of Cu atoms at Pb atomic sites, contracts the unit cell volume minutely, resulting in an enhanced repulsive interaction [1,2]. It is proposed that, $Cu^{2+}$ replacement of $Pb^{2+}$ generates stress, which is not relieved due to the structural uniqueness of LK-99 [1], unlike as in CuO and Fe based superconductor system, where stress is relieved due to structural freedom [16,17]. The strain induced superconductivity is also well known, be it from external factors (pressure and temperature) [4-8] or internal factors (doping) [16-19]. The external pressure or the doping induces strain in the structure of the material. The contraction due to $Cu^{2+}$ substitution of $Pb^{2+}$ (generating internal pseudo pressure) is given as plausible reason for superconductivity induction in parent Lead Apatite material [1-3].

In this article, we report the synthesis of LK-99 by following the method as given in [1,2], which involves the reaction of Lanarkite ($Pb_2SO_5$) with $Cu_3P$. The synthesized sample is been characterized for phase purity through XRD measurements. Our results, support the synthesis protocol of refs [1-3], to get LK-99, but do not approve its bulk superconductivity in terms of the room temperature magnetic levitation and bulk diamagnetism. The sample infact showed a paramagnetic behavior at 280K. We are en route to synthesize more LK-99 samples and will come up with our further findings.

**Synthesis and Experimental Details:**

For the synthesis of Lanarkite $Pb_2SO_5$, the required ingredients taken are $PbSO_4$ and PbO. The $PbSO_4$ is synthesized by the following reaction

$$Pb(NO_3)_2 + Na_2SO_4 = PbSO_4 + 2NaNO_3$$

The obtained white color powder is dried in oven and its phase purity is confirmed by performing the PXRD, as shown in the Fig. 1(a). The Lanarkite $Pb_2SO_5$ is prepared by mixing $PbSO_4$ and PbO and subject to the high temperature heat treatment. Here, we grow sample of $Pb_2SO_5$, as per same protocol being given ref. [1-2], by heating the ($PbSO_4$ and PbO) mixture an open-end alumina crucible, at 725ºC for 24 hours, the schematics are shown in inset of the Fig. 1(b). The other ingredient $Cu_3P$ is synthesized by reacting in the stoichiometric amount of Cu and P at 550ºC for 48 hours, as shown in inset of the Fig 1(c). The obtained powders of $Pb_2SO_5$ and $Cu_3P$ were taken in 1:1 stoichiometric ratio and heated to 925ºC for 10 hours to synthesize the final compound LK-99, i.e., $Pb_9Cu(PO_4)_6O$. The schematic of heat treatment for LK-99 is shown in inset of Fig. 2.

Subsequently, X-ray diffraction (XRD) measurements were performed on all finely crushed powder samples, which are matched with the JCPDS data. The obtained polycrystalline samples are subsequently, analyzed for their phase purity by XRD measurements using Rigaku made miniflex-II table top X-ray diffractometer equipped with Cu-$K_a$ radiation of 1.54 Å. The sample is examined for the presence of bulk superconductivity through magnetic levitation of permanent magnet on top of sample or the vice versa. To further elucidate on magnetic properties, the isothermal magnetization (MH) of the sample is done at 280K on MPMS SQUID magnetometer.

**Results and Discussions:**

Fig. 1(a) shows the PXRD (powder x-ray diffraction) of synthesized $PbSO_4$, all observed peaks match with the standard JCPDS data (36-1461) file and no extra peaks are seen, which confirms the purity of synthesized $PbSO_4$. Further we have synthesized the Lanarkite $Pb_2SO_5$ by solid state reaction method, following the procedure given in ref. [1,2]. As synthesized $Pb_2SO_5$ obtained after heat treatment is characterized by performing PXRD, again no additional peaks are observed, and the obtained pattern match with the JCPDS data file (72-1393). The PXRD pattern is shown in Fig. 1(b). Further, the other important ingredient for synthesizing the LK-99 required is $Cu_3P$, which is also synthesized by solid state reaction route. The PXRD of obtained $Cu_3P$ is shown in Fig. 1(c). The PXRD data matches with the JCPDS data (71-2261), confirming the bulk $Cu_3P$ phase. However, some extra peaks belonging to unreacted Cu are also observed, which are marked as # in the Fig. 1(c). The results of PXRD shown in Figs 1(a), (b) and (c) are very similar to that as reported in refs. [1-3].

After confirming, the formation of as synthesized $Cu_3P$ and $Pb_2SO_5$, both were taken in stoichiometric ratio of 1:1 as in ref. [1, 2] and vacuum sealed in quartz ampoule. After following heat treatment protocol (insets Fig. 2), the polycrystalline sample of LK-99 is obtained. This sample is named as LK-99. The PXRD of as obtained polycrystalline LK-99 samples is shown in Fig. 2, along with its heat treatment protocol in the inset of the same Fig. Besides main phase LK-99, some extra peaks of $Cu_2S$ are also seen, which are marked. Interestingly, the superconducting LK-99 being reported [1,2], also contain the minority $Cu_2S$ phase. Clearly our synthesized LK-99 is as good as being reported in ref. 1 and 2. Worth mentioning is the fact that the characteristic peak of Cu doped Lead Apatite in our PXRD at around 2(θ) of $18^0$ is of relatively much smaller intensity than in Ref.1. Also the amount of impurity $Cu_2S$ in our sample is relatively higher than as in ref. 1. The picture of the pieces of presently synthesized LK-99 sample are shown in Fig. 3. Further, our sample is not spongy as being reported in ref. 1. Rather it looks, as the same is slightly melted and reacted with quartz tube.

Now, let us move to the superconductivity characterization of the obtained LK-99. The very first test we did out of our curiosity was to check if a permanent magnet levitates over the obtained LK-99 sample. Fig. 4 shows the picture of tiny piece of sample sitting idle over a permanent magnet. This clearly shows that the obtained LK-99 sample is not superconducting. The isothermal magnetization (MH) plot taken on SQUID magnetometer at 280K, shows a clear paramagnetic behavior at 280K, see Fig. 5. This, corroborates the absence of magnetic levitation shown in Fig.4. Summarizing, Fig. 4 and 5, here at this point we are not been able to support the claims of room temperature superconductivity in LK-99 at ambient as being reported in ref. [1-3]. Worth mentioning is the fact that LK-99 is 1-D electrical material [1-3], hence appropriately doping of its Pb-apatite chains by Cu may be an uphill task.

In conclusion, our results, as of now do not approve the reported [1-3] appearance of bulk superconductivity in LK-99, in particular the room T levitation. We are en route of being synthesizing more samples of LK-99 and to further ascertain if room temperature ambient bulk superconductivity do exist in this compound. The authors feel that their timely taking up of the room temperature ambient bulk superconductivity of LK-99, with detailed synthesis protocol, will motivate various research groups to follow up the discovery of ref.1-3.

Authors would like to acknowledge the keen interest of Prof. Achanta Venu Gopal, Director CSIR-NPL in superconducting materials research. Dr. Pallavi Kushvaha is acknowledged for providing the MPMS based MH plot for our sample. The motivation and encouragement of Prof. G. Baskaran (IMSc/IITM) and Prof. D.D. Sarma (IISc) has been very instrumental in carrying out this research. The research is supported by in house project OLP-230232.

**Figure Captions:**

**Figure 1:** **(a)** PXRD of $PbSO_4$ synthesized by the chemical reaction. **(b)** PXRD of $Pb_2SO_5$ and the inset shows the schematic heat treatment diagram of $Pb_2SO_5$. **(c)** PXRD of synthesized $Cu_3P$ and the inset shows the schematic heat treatment diagram of $Cu_3P$.

**Figure 2:** PXRD of $925^0$C synthesized LK-99 sample, the inset shows the schematic heat treatment diagram for the same.

**Figure 3:** The picture of the pieces of $925^0$C synthesized LK-99 sample.

**Figure 4:** The picture of $925^0$C synthesized LK-99 sample sitting over the permanent magnet at room temperature.

**Figure 5:** The isothermal magnetization (MH) of studied 925 $^0$C synthesized LK-99 sample.

## References:

1. S. Lee, J. Kim, and Y.W. Kwon, arXiv: 2307.12008 (2023).
2. S. Lee, J. Kim, H.T. Kim, S. Im, S. An, K.H. Auh, arXiv: 2307.12037 (2023).
3. S. Lee, J. Kim, H.T. Kim, S. Im, S. An, K.H. Auh, J. Korean Cryst. Growth Cryst. Technol., **33**, 61 (2023).
4. A.P. Drozdov, M.I. Eremets, I. A. Troyan, V. Ksenofontov, and S. I. Shylin, S.I., Nature, **525**, 73 (2015).
5. A.P. Drozdov, P. P. Kong, V.S. Minkov, S.P. Besedin, M.A. Kuzovnikov, S. Mozaffari, L. Balicas, F.F. Balakirev, D.E. Graf, V.B. Prakapenka, and E. Greenberg, 2019. Nature, **569**, 7757 (2019).
6. M. Somayazulu, M. Ahart, A.K. Mishra, Z.M. Geballe, M. Baldini, Y. Meng, V.V. Struzhkin, and R.J. Hemley, Physical review letters, **122**(2), 027001 (2019).
7. G. Gao, L. Wang, M. Li, J. Zhang, R.T. Howie, E. Gregoryanz, V.V, Struzhkin, L. Wang, and S.T. John, Materials Today Physics, **21**, 100546 (2021).
8. A.D. Grockowiak, M. Ahart, T. Helm, W.A. Coniglio, R. Kumar, K. Glazyrin, G. Garbarino, Y. Meng, M. Oliff, V. Williams, and N.W. Ashcroft, Front. Elect Mat., **2**, 837651 (2022).
9. B. Lilia, R. Hennig, P. Hirschfeld, G. Profeta, A. Sanna, E. Zurek, W.E. Pickett, M. Amsler, R. Dias, M.I. Eremets, and C. Heil, Journal of Physics: Condensed Matter, **34**(18), 183002 (2022).
10. N. Dasenbrock-Gammon, E. Snider, R. McBride, H. Pasan, D. Durkee, N. Khalvashi-Sutter, S. Munasinghe, S.E. Dissanayake, K.V. Lawler, A. Salamat, and R.P. Dias, Nature **615**, 244 (2023).
11. D.P. Thapa, S. Islam, S.K. Saha, P.S. Mahapatra, B. Bhattacharyya, T.P. Sai, R. Mahadevu, S. Patil, A. Ghosh, and A. Pandey, arXiv:1807.08572 (2018).
12. V.P.S. Awana, J. Sup. Novel Mag. **31**, 3387 (2018).
13. V.P.S. Awana, I. Felner, S. Ovchinnikov, and J. W. A. Robinson J. Sup. Novel Mag. **36**, 1085 (2023).
14. J.E. Hirsch and F. Marsiglio, Phys. Rev. B. **103**, 134505 (2021).
15. Science Cast, https://www.sciencecast.org/casts/suc384jly50n
16. Q.Y. Wang, Z. Li, W.H. Zhang, Z.C. Zhang, J.S. Zhang, W. Li, H. Ding, Y.B. Ou, P. Deng, K. Chang, and J. Wen, Chinese Physics Letters, **29**(3), 037402 (2012).
17. Y. Zhong, Y. Wang, S. Han, Y.F. Lv, W.L. Wang, D. Zhang, H. Ding, Y.M. Zhang, L. Wang, K. He, and R. Zhong, Science Bulletin, **61**(16), 1239 (2016).
18. E.M. Choi, A. Di Bernardo, B. Zhu, P. Lu, H. Alpern, K.H. Zhang, T. Shapira, J. Feighan, X. Sun, J. Robinson, and Y. Paltiel, Science Advances, **5**(4), 5532 (2019).
19. J. Engelmann, V. Grinenko, P. Chekhonin, W. Skrotzki, D.V. Efremov, S. Oswald, K. Iida, R. Hühne, J. Hänisch, M. Hoffmann, and F. Kurth, Nature communications, **4**(1), 2877 (2013).

Fig. 1(a):

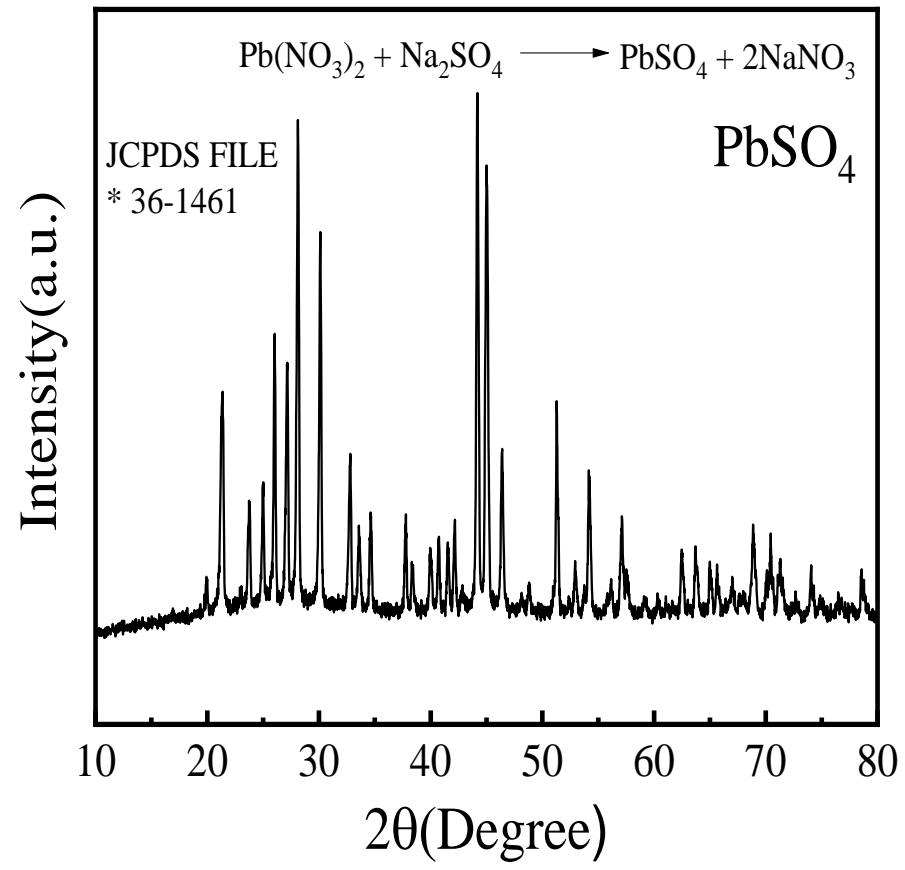

Fig. 1(b):

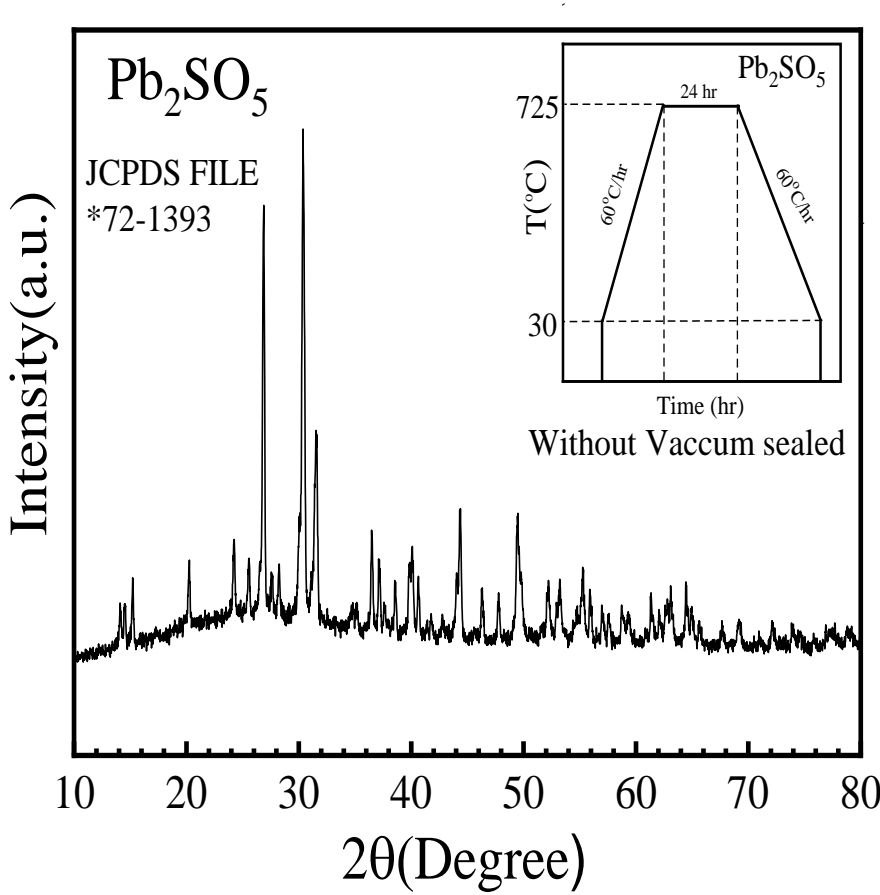

Fig. 1(c):

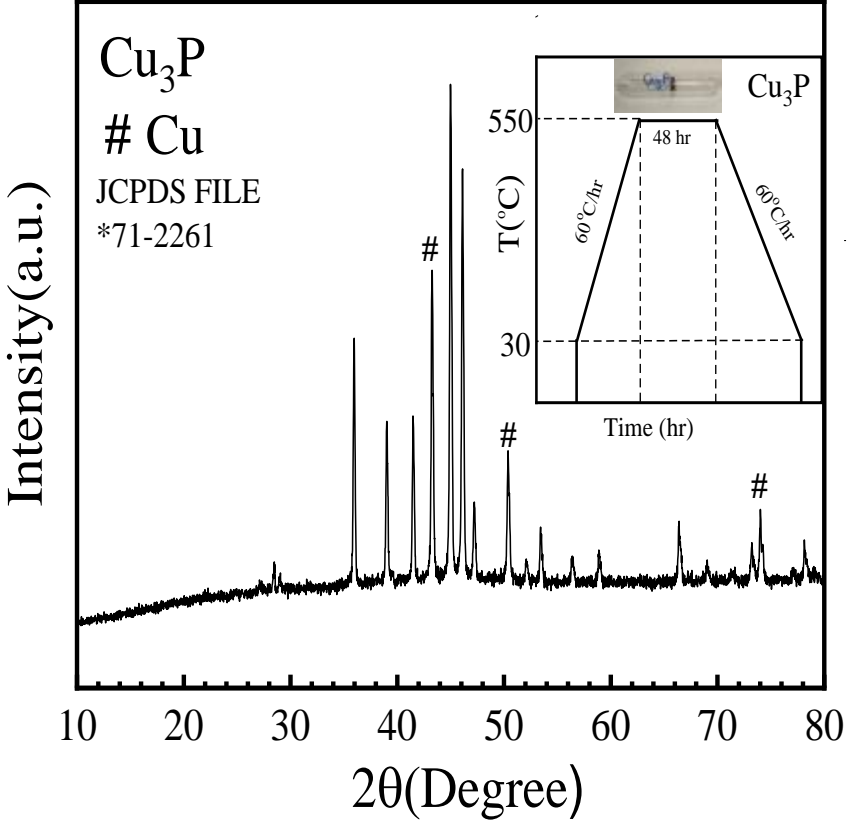

Fig. 2:

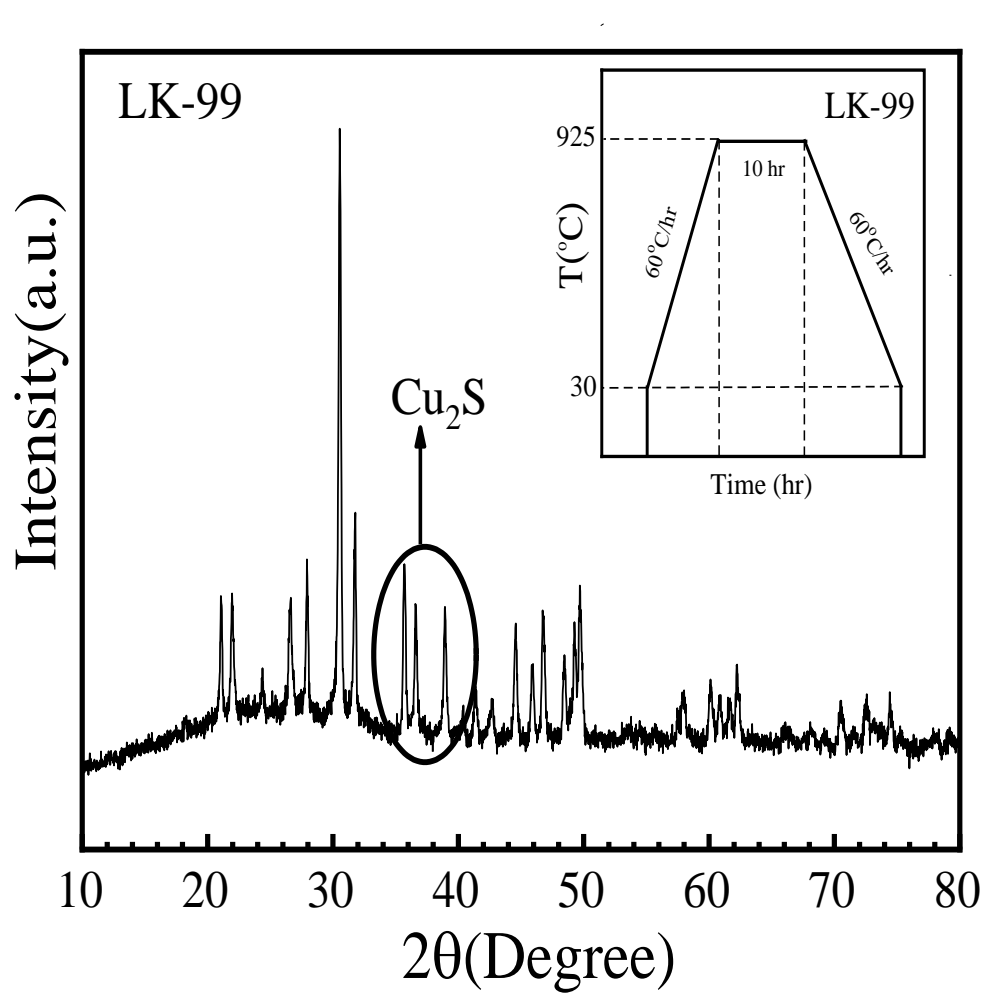

Fig. 3

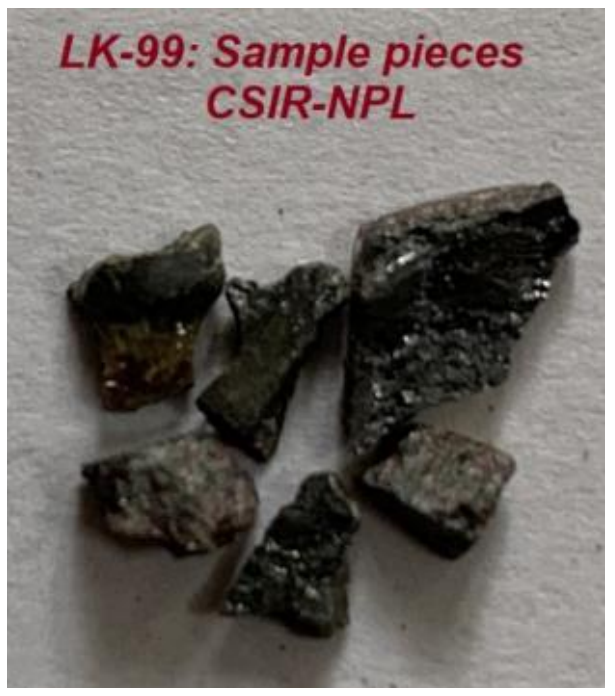

Fig. 4

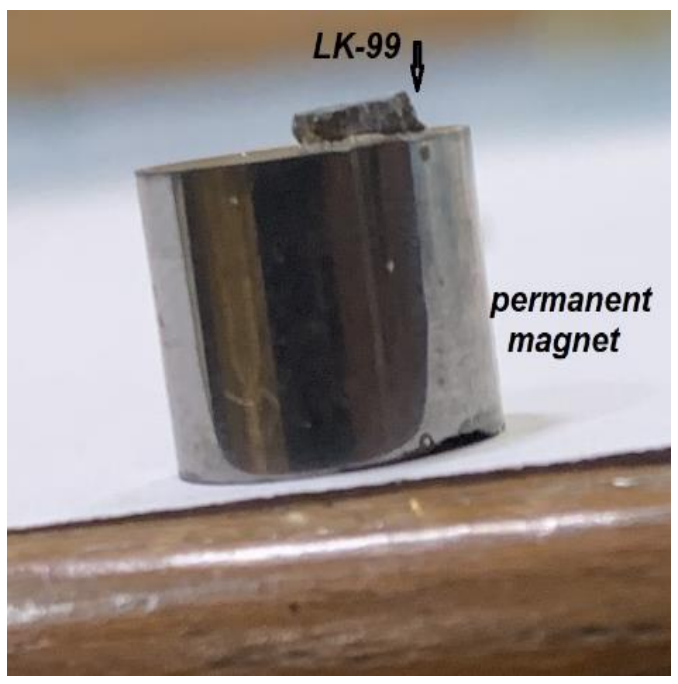

Fig. 5

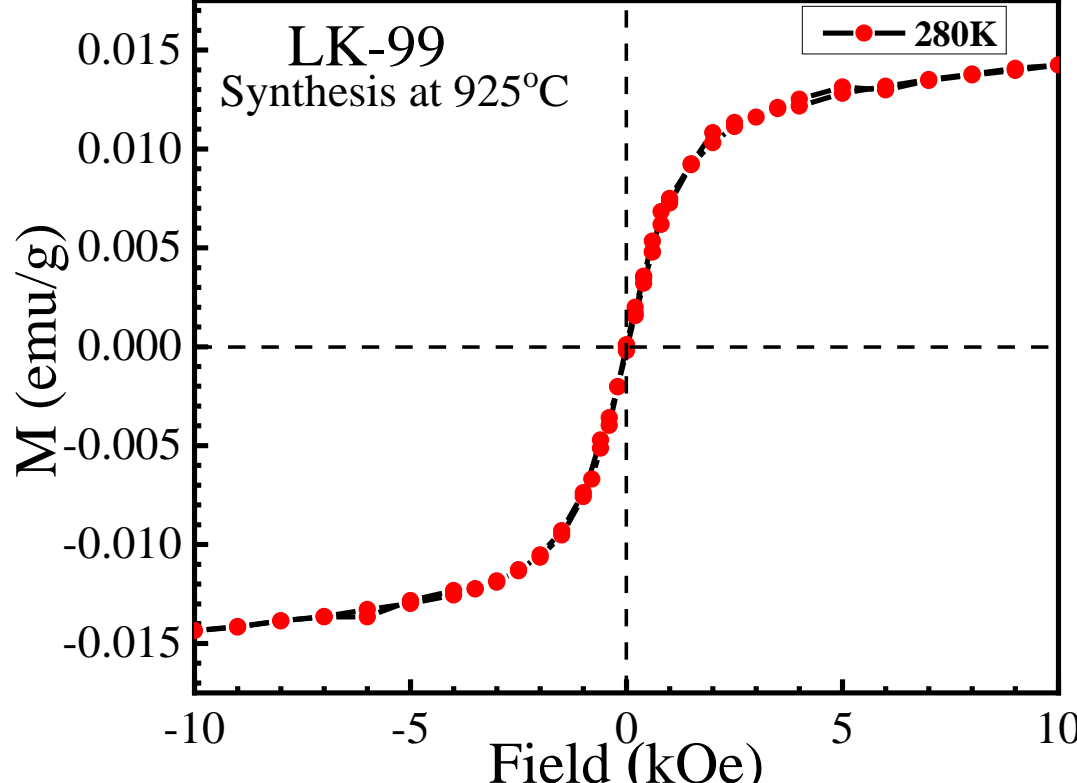